\documentclass[a4paper, 12pt]{article}

\usepackage[english]{babel}
\usepackage[sort&compress]{natbib}
\bibpunct{(}{)}{;}{a}{}{,} 

\usepackage{amsthm, amsmath, amssymb, mathrsfs, multirow, url, subfigure}
\usepackage{graphicx} 
\usepackage{ifthen} 
\usepackage{amsfonts}
\usepackage[usenames]{color}
\usepackage{fullpage}
\usepackage{booktabs}

\theoremstyle{plain}

\theoremstyle{definition}

\theoremstyle{remark}

\newcommand{\E}{\mathsf{E}}

\newcommand{\nm}{{\sf N}}

\newcommand{\rad}{{\sf Rad}}

\newcommand{\RR}{\mathbb{R}}

\newcommand{\PP}{\mathbb{P}}

\newcommand{\eps}{\varepsilon}

\newcommand{\model}{\mathscr{P}}

\usepackage{stackrel}

\newcommand{\trace}{\mathrm{tr}}

\title{A comparison of learning rate selection methods in generalized Bayesian inference}
\author{Pei-Shien Wu\footnote{Department of Statistics, North Carolina State University; {\tt pwu9@ncsu.edu}, {\tt rgmarti3@ncsu.edu}} \quad and \quad Ryan Martin$^*$}
\date{\today}

\begin{document}

\maketitle 

\begin{abstract}    
Generalized Bayes posterior distributions are formed by putting a fractional power on the likelihood before combining with the prior via Bayes's formula.  This fractional power, which is often viewed as a remedy for potential model misspecification bias, is called the {\em learning rate}, and a number of data-driven learning rate selection methods have been proposed in the recent literature.  Each of these proposals has a different focus, a different target they aim to achieve, which makes them difficult to compare.  In this paper, we provide a direct head-to-head comparison of these learning rate selection methods in various misspecified model scenarios, in terms of several relevant metrics, in particular, coverage probability of the generalized Bayes credible regions.  In some examples all the methods perform well, while in others the misspecification is too severe to be overcome, but we find that the so-called generalized posterior calibration algorithm tends to outperform the others in terms of credible region coverage probability. 

\smallskip

\end{abstract}

\section{Introduction}
\label{S:intro}  

Specification of a sound model is a critical part of an effective statistical analysis. This is especially true for a Bayesian approach, since the statistical model or likelihood is explicitly used to construct the posterior distribution from which inferences will be drawn.  However, it is common in applications to know relatively little about the phenomenon under investigation, which impacts our ability to specify a sound statistical model. 
For this reason, the effects of model misspecification have received considerable attention; in the Bayesian literature, this includes  \citet{berk1966limiting}, \citet{bunke1998asymptotic}, \citet{diaconis1986consistency, diaconis1986inconsistent}, \citet{kleijn2006misspecification, kleijn2012bernstein}, \citet{walker2013}, \citet{deblasi.walker.2013}, \citet{rvr.sriram.martin}, and \citet{grunwald.ommen.scaling}.  In the most general case, misspecification implies that there is no ``true'' parameter value that the posterior could concentrate around.  Instead, under suitable conditions, the posterior will concentrate around a ``best'' parameter value, one that minimizes the Kullback--Leibler divergence of the posited model from the true data-generating distribution.  But even in those relatively nice cases, where the best parameter value around which the posterior concentrates could be meaningful, or even equal to the real quantity of interest, there is reason for concern.  \citet{kleijn2012bernstein} showed that misspecification can also affect the posterior spread, which means that the actual frequentist coverage probability of the Bayesian posterior credible region can be arbitrarily far below the advertised/nominal level.  

Real coverage probabilities differing significantly from advertised levels is a serious concern \citep{fraser2011, martin.nonadditive}.
A gap between real and advertised coverage probabilities can can have various causes, but here we focus on model misspecification.  To avoid this misspecification bias, there are a few options: first, to take an approach that does not depend explicitly on a statistical model; second, to work with a model that is sufficiently broad that misspecification is virtually impossible; and third, to make some adjustments to correct for potential model misspecification. From a Bayesian perspective, the first fix is not available, since Bayes's formula requires a likelihood function. The second fix amounts to the use of Bayesian nonparametrics but, when the quantity of interest is a low-dimensional feature of the full distribution, introducing an infinite-dimensional parameter, with the computational and statistical challenges that entails, would be overkill.  This leaves only the third option, but what kind of adjustments might the Bayesian consider?  Recently, \citet{grunwald.ommen.scaling} argued that a certain adjustment to the usual Bayesian posterior distribution could repair inconsistencies resulting from model misspecification.  The goal of the present paper is to investigate the extent to which Gr\"unwald and van Ommen's proposed adjustment---and other related adjustments in the literature---can close the gap between the real and advertised coverage probabilities of Bayesian posterior credible regions affected by model misspecification.  

More specifically, here we will be working in the so-called {\em generalized Bayes} framework, which differs from the traditional Bayes framework only in that a {\em learning rate} parameter, a power $\eta > 0$ on the likelihood function, is introduced.  That is, if we have data $D^n$ and a posited statistical model $P_\theta^n$, indexed by a parameter $\theta$ in $\Theta$, then the generalized Bayes posterior distribution for $\theta$ is 
\begin{equation}
\label{eq:gbayes}
\Pi_n^{(\eta)}(d\theta) \propto L_n^\eta(\theta) \, \Pi(d\theta), \quad \theta \in \Theta, 
\end{equation}
where $\theta \mapsto L_n(\theta) = L(\theta; D^n)$ is the likelihood function and $\Pi$ is a prior distribution on $\Theta$.   Among the first papers to adopt such an approach is \citet{walker.hjort.2001}, followed up on by \citet{zhang2006a}. \citet{bissiri.holmes.walker.2016} showed that \eqref{eq:gbayes} is the principled way to update prior beliefs when the model is potentially misspecified, and that the appearance of a non-trivial learning rate is a necessary by-product.  A different connection between robustness and learning rate $\eta < 1$ was made recently in \citet{miller.dunson.power}.  More details about model misspecification and generalized Bayes are given in Section~\ref{S:background}.  

Gr\"unwald and van Ommen's claim is that, for a sufficiently small learning rate $\eta$, certain model misspecification biases can be repaired.  Of course, the threshold defining ``sufficiently small'' cannot be known in practice, so some data-driven choices are required.  \citet{grunwald2012, grunwald.safe}, \citet{grunwald.ommen.scaling}, and \citet{heide2020safe} developed a so-called {\em SafeBayes} algorithm to choose the learning rate $\eta$, based on minimizing a sequential risk measure.  A number of other learning rate selection methods have been proposed recently, including the two distinct information matching strategies in \citet{holmes2017assigning} and \citet{lyddon2019general}, and bootstrap-motivated calibration methods of \citet{syring2019calibrating}. 
Since the role played by the learning rate is relatively unfamiliar and since the various methods differ significantly in terms of their motivations and implementations, it would be beneficial to see a head-to-head comparison in terms of some standard metrics, for example, the validity and efficiency of the corresponding generalized Bayes credible regions.  This paper aims to fill this gap.  

The remainder of this paper is organized as follows. In Section~2, we discuss the behavior of Bayesian posterior distribution under a misspecified model and define and review the literature on generalized Bayes posteriors.  For the latter, the choice of learning rate is essential, and we provide details for four recently proposed learning rate selection methods in Section~3.  Then, in Section~4, we show a simple illustrative example to give some intuition about how the different methods perform and, in particular, this suggests that the methods which are not designed specifically to calibrate the credible region's coverage probability may not be able to achieve the nominal level in general.  Simulation results are presented in Sections~5--6, for linear and binary regression models, and the take-away message is that the method of \citet{syring2019calibrating} is more stable in achieving the coverage probability across different sample sizes and misspecification degrees compared to the others. Some concluding remarks are given in Section~7.

\section{Background}
\label{S:background}

\subsection{Model misspecification}
\label{SS:misspecified}

Suppose we have data $D^n$ which, for simplicity, we assume consists of independent and identically distributed observations: either response variables $Y_i$ only or predictor and response variables pairs $(X_i,Y_i)$, $i=1,\ldots,n$.  To analyze these data, we posit a statistical model $\model = \{P_\theta: \theta \in \Theta\}$, a collection of probability measures on the sample space, indexed by a parameter $\theta$ taking values in the parameter space $\Theta$.  From this model and the observed $D^n$, we obtain a likelihood function $L_n$.  The likelihood summarizes the information in the data relative to the posited model, which can be combined with prior information encoded in a distribution $\Pi$ for $\theta$ on $\Theta$ via Bayes's formula:
\[ \Pi_n(d\theta) \propto L_n(\theta) \, \Pi(d\theta), \quad \theta \in \Theta. \]
In the Bayesian paradigm, inferences about $\theta$ are drawn based on the posterior distribution $\Pi_n$, e.g., degrees of belief about the truthfulness of an assertion ``$\theta \in A$,'' for $A \subset \Theta$, are summarized by the posterior probability $\Pi_n(A)$.   

Let $P^\star$ denote the true distribution of $Y_1$ or of $(X_1,Y_1)$.  If the model is correctly specified, then there exists a $\theta^\star \in \Theta$ such that $P^\star = P_{\theta^\star}$.  In that case, under suitable regularity conditions, inference based on the posterior distribution will be valid, at least asymptotically.  That is, $\Pi_n$ will concentrate its mass around $\theta$ as $n \to \infty$ and, moreover, the Bernstein--von Mises theorem \citep[e.g.,][Theorem~??]{van2000asymptotic}, states that $\Pi_n$ is approximately a normal distribution, centered at the maximum likelihood estimator $\hat\theta_n$, with covariance matrix proportional to the inverse of the Fisher information matrix at $\theta^\star$.  This implies, among other things, that credible regions derived from $\Pi_n$ closely resemble those asymptotic confidence regions based on likelihood theory.  Therefore, asymptotically, the Bayesian posterior credible regions will have frequentist coverage probability close to the advertised level.  

If the model is incorrectly specified, in the sense that $P^\star \not\in \model$, then there are several challenges.  First, there is no ``true'' $\theta^\star$, which creates some challenges in interpretation.  Indeed, the maximum likelihood estimator, Bayes posterior, or any other model-based procedure will identify the {\em Kullback--Leibler projection} of $P^\star$ onto the model, i.e., 
\[ \theta^\dagger = \arg\min_\theta K(P^\star, P_\theta), \]
where $K(P^\star, P_\theta) = \int \log(dP^\star / dP_\theta) \, dP^\star$ is the Kullback--Leibler divergence of $P_\theta$ from $P^\star$.
In general, $\theta^\dagger$ does not have a real-life interpretation but, in some cases, certain features of $P^\star$ can be identified based on a misspecified model.  For example, if $\model$ is an exponential family, then the mean function of the exponential family model, evaluated at $\theta^\dagger$, equals the mean of $P^\star$ \cite[][Example~2]{bunke1998asymptotic}.  Another similar case is considered in Section~\ref{S:linear}.  The second challenge is that, even in the case where $\theta^\dagger$ has a (limited) real-life interpretation, misspecification can still negatively impact  posterior inferences.  \citet{kleijn2012bernstein} established a version of the Bernstein--von Mises theorem under model misspecification which states that, under certain regularity conditions, the posterior $\Pi_n$ will be approximately normal, with mean equal to the maximum likelihood (or M-) estimator $\hat\theta$ and covariance matrix $V_{\theta^\dagger}^{-1}$, where 
\[ V_{\theta^\dagger} = \int \Bigl( \frac{\partial^2 \log p_\theta}{\partial\theta \partial \theta^\top} \Bigr|_{\theta=\theta^\dagger} \Bigr) \, dP^\star, \]
and $p_\theta$ is the density function corresponding to $P_\theta$.  The problem, of course, is that $V_{\theta^\dagger}^{-1}$ is {\em not} the asymptotic covariance matrix of $\hat\theta_n$; the latter, as shown by \citet{huber1967behavior} and \citet{van2000asymptotic}, has the famous sandwich formula $V_{\theta^\dagger}^{-1} \Lambda_{\theta^\dagger} V_{\theta^\dagger}^{-1}$, where 
\[ \Lambda_{\theta^\dagger} = \int \Bigl( \frac{\partial \log p_\theta}{\partial\theta} \Bigr|_{\theta=\theta^\dagger} \Bigr) \Bigl( \frac{\partial \log p_\theta}{\partial\theta} \Bigr|_{\theta=\theta^\dagger} \Bigr)^\top \, dP^\star. \]
The implication of this covariance mismatch is that, even if the quantity of interest can be identified under the misspecified model, the frequentist coverage probability of the Bayes posterior credible sets could be arbitrarily far from the advertised level.  The question is: {\em can something be done to correct this problematic behavior?}

\subsection{Generalized Bayes}

Modifying the usual Bayesian update with a learning rate $\eta$ as in \eqref{eq:gbayes} is a simple change, but it has some unexpected consequences.  In particular, \citet{walker.hjort.2001} showed that, for a correctly specified model, consistency of the generalized Bayes posterior $\Pi_n^{(\eta)}$ in \eqref{eq:gbayes} could be established for any $\eta < 1$, with only local conditions on the prior---as opposed to the local and global conditions required for consistency with $\eta=1$ \citep[e.g.,][]{ggr1999, bsw1999}.  The intuition given by \citet{walker.lijoi.prunster.2005a} is that inconsistencies result from the posterior over-fitting or tracking the data too closely, and the fractional power discounts the data slight to prevent this over-fitting.  The Walker--Hjort result has been extended to cover posterior concentration rates, where the removal of the global prior conditions---usually formulated in terms of metric entropy \citep[cf.][]{ggv2000, ghosal.vaart.book}---leads to simpler proofs and generally (at least slightly) faster rates.  See \citet{zhang2006a} for one of the first papers exploring these ideas, and \citet{bhat.pati.yang.fractional} and \citet{grunwald.mehta.rates} for more recent contributions.  The fractional power has also been employed recently in work on high-dimensional problems using an empirical or data-driven prior \citep[e.g.,][]{martin.walker.deb, martin.mess.walker.eb, ebpred} where, again, the fractional power is motivated by the desire to prevent over-fitting; see, also, \citet{martin.horseshoe.discuss} and \citet{ebcvg} for some potential benefits of $\eta < 1$ to uncertainty quantification.  

When the model is misspecified, however, the learning rate is less about convenience and more about necessity.  \citet{bissiri.holmes.walker.2016} showed that the generalized Bayes update \eqref{eq:gbayes} is fundamental from a decision-theoretic point of view.  Moreover, they argue that the learning rate $\eta$ naturally emerges since, roughly, the parameter $\theta^\dagger$ being estimated is defined by minimizing the expectation of a loss function $\theta \mapsto \log p_\theta$, and since that minimization problem is invariant to scalar multiples of the loss, the learning rate should appear in the posterior \eqref{eq:gbayes}. In fact, the loss function interpretation makes their result much more general.  In many cases, it is more natural to formulate the inference problem with a loss function rather than a statistical model.  These are often referred to as {\em Gibbs posterior distributions}; see  \citet{syring.martin.mcid, syring2019calibrating, syring.martin.image, gibbs.general}, \citet{gibbs.quantile}, and \citet{wang.martin.auc}. 

Beyond recognizing the importance of the learning rate parameter, an actual value for $\eta$ needs to be set in practical applications.  Several recent papers---including \citet{grunwald.ommen.scaling}, \citet{holmes2017assigning}, \citet{lyddon2019general}, and \citet{syring2019calibrating}---have proposed data-driven choices for the learning rate, with different motivations.  Section~\ref{S:selection} describes these methods.  The remainder of the paper is focused on a comparison of these different learning rate methods.  



\section{Learning rate selection methods}
\label{S:selection}

\subsection{Gr\"unwald's SafeBayes}

\citet{grunwald.ommen.scaling} observe that, when the model is non-convex and misspecified, there is a chance for {\em hyper-compression}.  This is the term they use to describe the seemingly paradoxical result that the Bayesian predictive distribution can be closer, in a Kullback--Leibler sense, to the true $P^\star$ then the within-model Kullback--Leibler minimizer $P_{\theta^\dagger}$.  What makes this possible is non-convexity: the predictive distribution is an average of in-model distributions $P_\theta$ which, without convexity, could end up outside the model and potentially closer to $P^\star$ than is $P_{\theta^\dagger}$.  Besides being counter-intuitive, hyper-compression also reveals a practical problem, namely, inconsistency---that the posterior distribution is not concentrating its mass near $\theta^\dagger$ as expected.  To overcome this, \citet{grunwald.ommen.scaling} suggest to work with a new  (hypothetical) model, with densities 
\[ p_\theta^{(\eta)}(x,y) = p^\star(x,y) \bigl\{ p_\theta(y \mid x) / p_{\theta^\dagger}(y \mid x) \bigr\}^\eta, \]
indexed by a parameter $\eta > 0$.  We say this model is ``hypothetical'' because it depends on $p^\star$ and $\theta^\dagger$, two ingredients that are not available to the data analyst. However, {\em if $\eta$ is sufficiently small}, in the sense that $\int p_\theta^{(\eta)}(x,y) \, dx \, dy$ is strictly less than 1, then this indeed defines a genuine statistical model, with two interesting properties:
\begin{itemize}
\item it is not misspecified, i.e., the Kullback--Leibler minimizer is $\theta^\dagger$ and $p_{\theta^\dagger}^{(\eta)} = p^\star$; 
\vspace{-2mm}
\item and the Bayesian posterior based on this new model is precisely the generalized Bayes posterior $\Pi_n^{(\eta)}$, with learning rate $\eta$, as in \eqref{eq:gbayes}.  
\end{itemize}
Since this new model is not misspecified, hyper-compression and inconsistency of the $\eta$-generalized Bayes posterior can be avoided.  So: {\em how to choose $\eta$ sufficiently small?}

\citet{grunwald.ommen.scaling}, building on work in, e.g.,  \citet{grunwald2012}, argue that the so-called {\em SafeBayes} algorithm will select a learning rate $\eta$ that is sufficiently small in the sense above.  Define the cumulative expected log-loss under the $\eta$-generalized Bayes posterior distribution, as a function of $\eta$:
\[ \eta \mapsto \sum_{i=1}^n \int -\log p_\theta(Y_i \mid X_i) \, \Pi_{i-1}^{(\eta)}(d\theta). \]
The SafeBayes algorithm returns the minimizer, $\hat\eta$, of this function over the range $\eta \in [0,1]$.  \citet{grunwald2012} presents an argument for why the SafeBayes choice of $\hat\eta$ works in the sense of being sufficiently small as in the discussion above.   

What we have described here is one of two versions of the SafeBayes algorithm presented in \citet{grunwald.ommen.scaling}, namely, the ``R-SafeBayes'' version.  In our examples below, we found that the ``R'' version outperformed the other---namely, the ``I-SafeBayes'' version---so here we only discuss the former.

\subsection{\citet{holmes2017assigning}}

Following \citet{bissiri.holmes.walker.2016}, the Bayesian and generalized Bayesian frameworks can be considered simply as rules for using data to update prior beliefs to posterior beliefs.  As such, it makes sense to consider how much information has been gained from the update, by comparing the prior to the posterior.  Of course, this information gain depends on both the updating rule and on the data, and \citet{holmes2017assigning} proposed a procedure for selecting the learning rate $\eta$ based on matching the expected information gain between Bayes and generalized Bayes updates.  

More formally, if $I_\eta(x,y)$ denotes the information gain in the generalized Bayes update from prior to posterior based on learning rate $\eta$ and data values $(x,y)$, then \citet{holmes2017assigning} propose to set $\eta$ such that 
\begin{equation}
\label{eq:info.match}
\int I_\eta(x,y) \, P^\star(dx,dy) = \int I_1(x,y) \, P_{\theta^\dagger}(dx,dy), 
\end{equation}
where $I_1(\cdot)$ denotes the information gain in the standard Bayesian update.  The specific choice of information measure they recommend is the {\em Fisher divergence} 
\[ I_\eta(x,y) = \int \{ \nabla \log \pi_{x,y}^{(\eta)}(\theta) - \nabla \log \pi(\theta) \}^2 \, \pi(\theta) \, d\theta, \]
where $\pi_{x,y}^{(\eta)}$ denotes the generalized Bayes posterior based on data $(x,y)$ and learning rate $\eta$, and $\nabla$ is the gradient operator.  Then it is straightforward to check that $I_\eta(x,y) = \eta^2 I_1(x,y)$ and, therefore, by \eqref{eq:info.match}, an ``oracle'' learning rate is given by 
\[ \eta^\star = \Bigl\{ \frac{\int I_1(x,y) \, P_{\theta^\dagger}(dx,dy)}{\int I_1(x,y) \, P^\star(dx,dy)} \Bigr\}^{1/2}.  \]
Of course, both $P^\star$ and $P_{\theta^\dagger}$ are unknown, so $\eta^\star$ cannot be evaluated, but the expectations can be estimated with the actual data $\{(X_i,Y_i): i=1,\ldots,n\}$.  That is, 
\[ \hat\eta = \Bigl\{ \frac{\int I_1(x,y) \, P_{\hat\theta_n}(dx,dy)}{\int I_1(x,y) \, \PP_n(dx,dy)} \Bigr\}^{1/2},  \]
where $\hat\theta_n$ is the maximum likelihood estimator of $\theta$ under the model---which is an estimate of $\theta^\dagger$---and $\PP_n$ is the empirical distribution of the data.


\subsection{\citet{lyddon2019general}}

The learning rate selection strategy presented in \citet{lyddon2019general} is motivated by the weighted likelihood bootstrap approach of \citet{newton1994approximate}, which was shown to generate bootstrap samples that have the same asymptotic distribution as Bayesian posterior distribution under a correctly specified model. For the case of a misspecified model, \citet{lyddon2019general} proposed a modified the weighted likelihood bootstrap approach which replaces the ordinary bootstrap with the Bayesian bootstrap, and establish its asymptotic limiting distribution.  Then following a strategy similar to that in \citet{holmes2017assigning} described above, they propose to choose $\eta$ in order to match the limiting $\eta$-generalized Bayes posterior to that of this modified likelihood bootstrap.  They then show that, using the notation defined at the end of Section~\ref{SS:misspecified}, an ``oracle'' learning rate is 
\[ \eta^\star = \frac{\trace(V_{\theta^\dagger} \Lambda_{\theta^\dagger}^{-1} V_{\theta^\dagger})}{\trace(V_{\theta^\dagger})}. \]
Again, since $\theta^\dagger$ and $P^\star$ are unknown, this oracle value cannot be evaluated.  However, a data-driven choice $\hat\eta$ can be obtained by replacing $\theta^\dagger$ the maximum likelihood estimator and the expectations with respect to $P^\star$ in $V_\theta$ and $\Lambda_\theta$, respectively, with expectations with respect to the empirical distribution $\PP_n$.

\subsection{\citet{syring2019calibrating}}

The three previous subsections describe principled learning rate selection strategies, but none of those are tailored so that the generalized posterior distribution achieves any specific and desirable properties.  Since the learning rate's effect on the posterior is to control the spread, \citet{syring2019calibrating} proposed to tune the learning rate such that posterior credible sets approximately achieve the nominal frequentist coverage probability.  

The coverage probability function is given by 
\[ c_\alpha(\eta \mid P^\star) = P^\star\{ C_\alpha^{(\eta)}(D^n) \ni \theta^\dagger\}, \]
where $C_\alpha^{(\eta)}$ is the $\eta$-generalized Bayes $100(1-\alpha)$\% credible region for $\theta$, e.g., a highest posterior density region, $\theta^\dagger$ is the Kullback--Leibler minimizer in the model, treated as a functional of $P^\star$, and $D^n = \{(X_i,Y_i): i=1,\ldots,n\}$ is the iid data set from $P^\star$.  Then the goal is to find $\eta$ such that $c_\alpha(\eta \mid P^\star) = 1-\alpha$.  Of course, lots of the quantities involved in this equation are unknown, but they can be estimated.  In particular, if $P^\star$ is replaced by the empirical distribution $\PP_n$, then the new equation is 
\[ c_\alpha(\eta \mid \PP_n) := \PP_n\{ C_\alpha^{(\eta)}(D^n) \ni \hat\theta_n\} = 1-\alpha, \]
where $\hat\theta_n$ is the maximum likelihood estimator based on the observed data, i.e., the ``$\theta^\dagger$-functional'' applied to $\PP_n$.  Even this alternative coverage probability function cannot be evaluated, since it requires enumerating all $n^n$ possible with-replacement samples from the observed data, but a bootstrap approximation is possible.  That is, for $B$ bootstrap samples $\tilde D_1^n,\ldots,\tilde D_B^n$, calculate 
\[ \hat c_\alpha(\eta \mid \PP_n) = \frac1B \sum_{b=1}^B 1\{ C_\alpha^{(\eta)}(\tilde D_b^n) \ni \hat\theta_n\}. \]
To solve the equation, $\hat c_\alpha(\eta \mid \PP_n) = 1-\alpha$, \citet{syring2019calibrating} recommend a stochastic approximation scheme that iteratively defines a learning rate sequence $(\eta_t)$ as
\[ \eta_{t+1} = \eta_t + k_t \{ \hat c_\alpha(\eta \mid \PP_n) - (1-\alpha)\}, \quad t \geq 1, \]
where $k_t$ is a sequence such that $\sum_t k_t = \infty$ and $\sum_t k_t^2 < \infty$.  When the $\eta_t$ sequence effectively converges, the limit is the suggested learning rate $\hat\eta$.  This is what \citet{syring2019calibrating} refer to as the generalized posterior calibration (GPC) algorithm.  Like SafeBayes, it is relatively expensive computationally---these algorithms require posterior computations for multiple learning rates and data sets---but the benefit is having a posterior distribution with meaningful spread, even in finite samples.

\section{Learning rates in a toy example}
\label{S:main}

Before we consider the effect of different learning rate selection methods in some non-trivial real-world problems, it helps to consider a simple example, one where some of the calculations can be done by hand, to develop some intuition about what to expect.  

Suppose that the posited model for iid data $Y^n=(Y_1,\ldots,Y_n)$ is $P_\theta = \nm(\theta, \sigma^2)$, with $\sigma > 0$ fixed, but that the true distribution is $P^\star = \nm(\theta^\star, \sigma^{\star 2})$, where $\sigma^\star > 0$ is potentially different from $\sigma$.  It is easy to confirm that the Kullback--Leibler minimizer satisfies $\theta^\dagger = \theta^\star$, but the misspecified variance can still cause problems, as we now demonstrate.  

It is intuitively clear that the generalized Bayes framework could completely resolve the model misspecification if the learning rate was chosen as $\eta^\star = (\sigma/\sigma^\star)^2$.  More formally, \citet{heide2020safe} show that, if the learning rate is no larger than this ratio, the generalized Bayes posterior will enjoy fast root-$n$ rate convergence properties, while \citet{syring2019calibrating} argue that learning rate equal to the ratio in order to achieve exact coverage of credible sets.  In any case, of course, one cannot make this learning rate choice in practice because it depends on the unknown value of true variance.  But this intuition tells us what the different learning rate selection methods' target should be.

\begin{figure}[t]
\begin{center}
\subfigure[$n=100$]{\scalebox{0.45}{\includegraphics{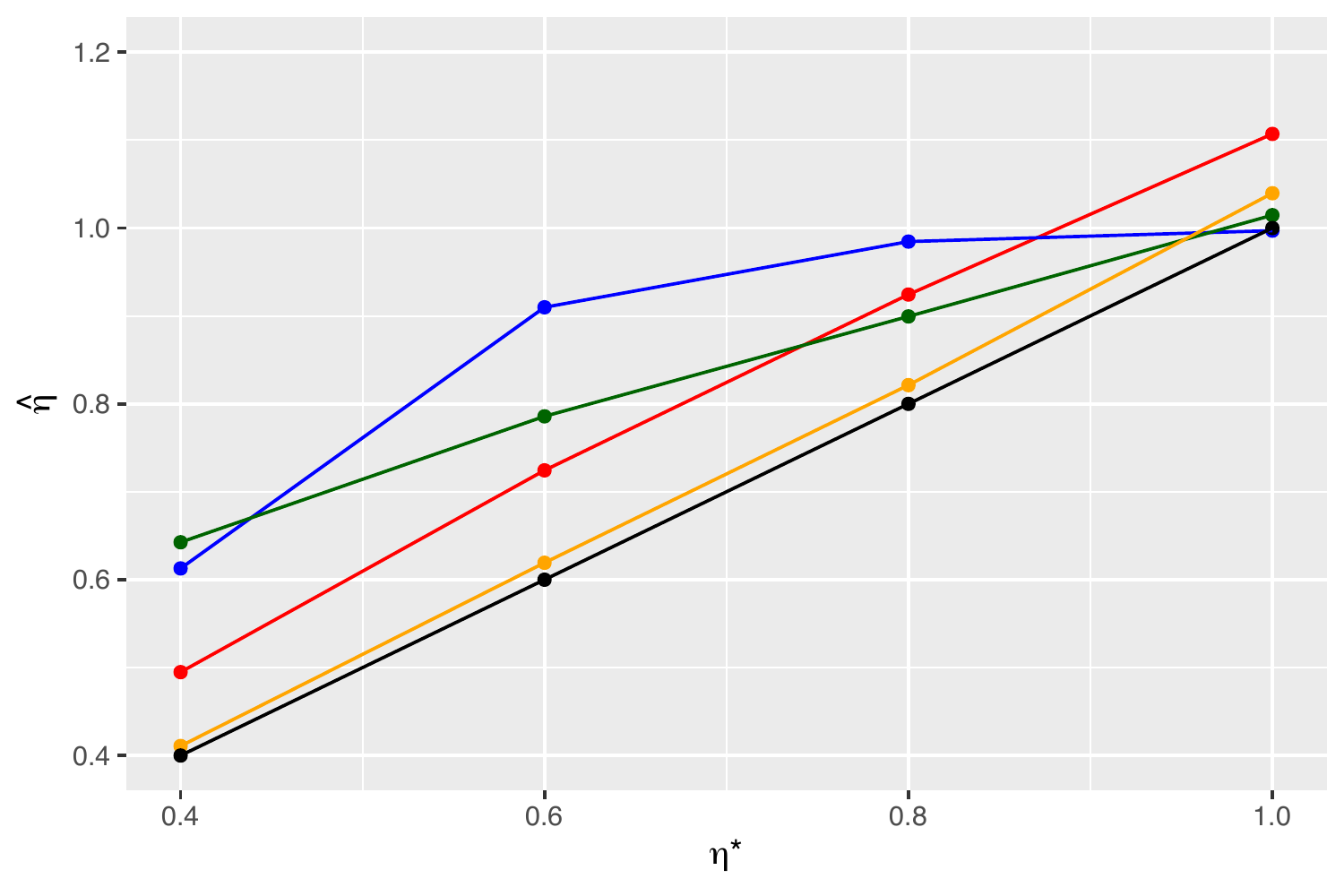}}}
\subfigure[$n=200$]{\scalebox{0.45}{\includegraphics{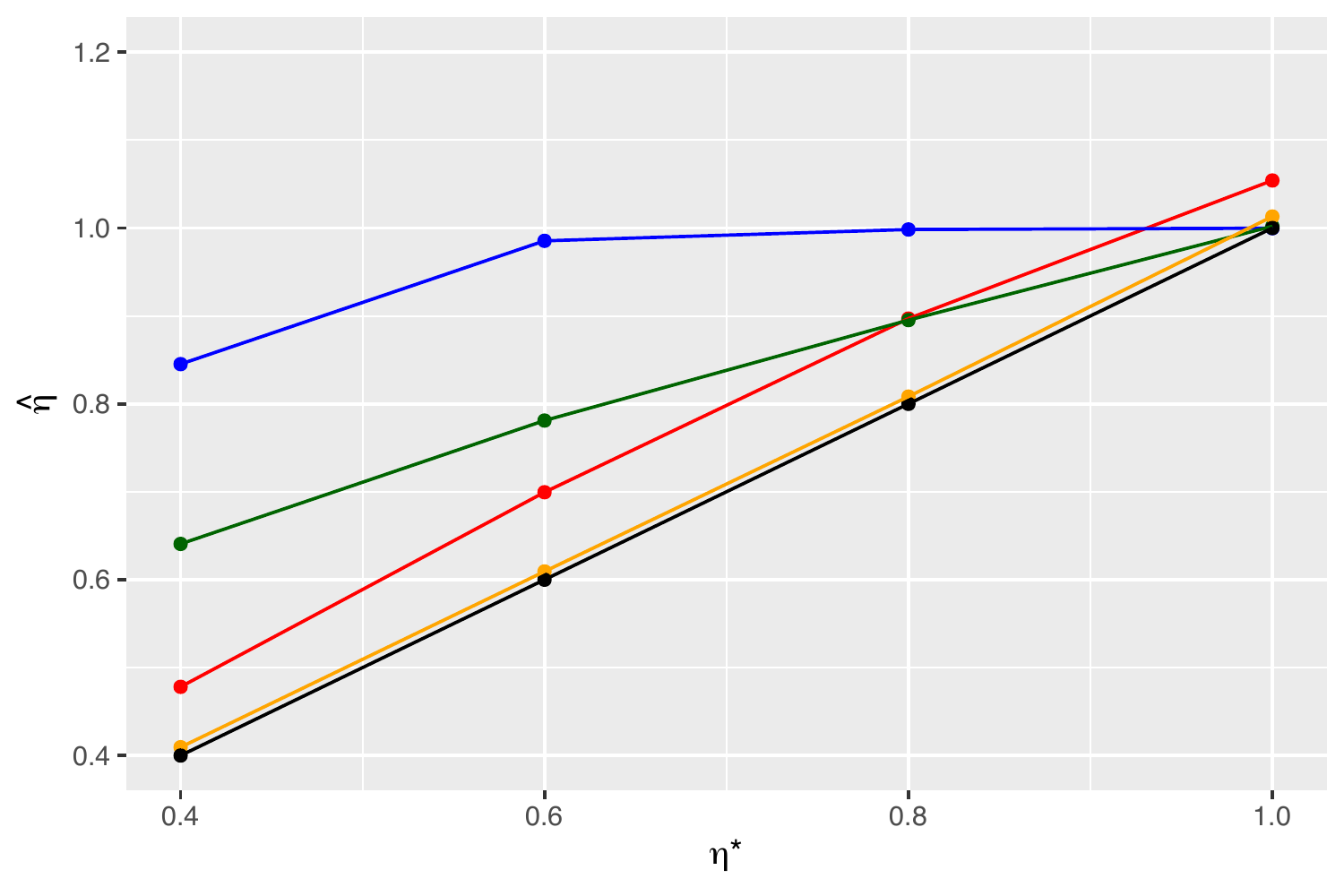}}}
\subfigure[$n=400$]{\scalebox{0.45}{\includegraphics{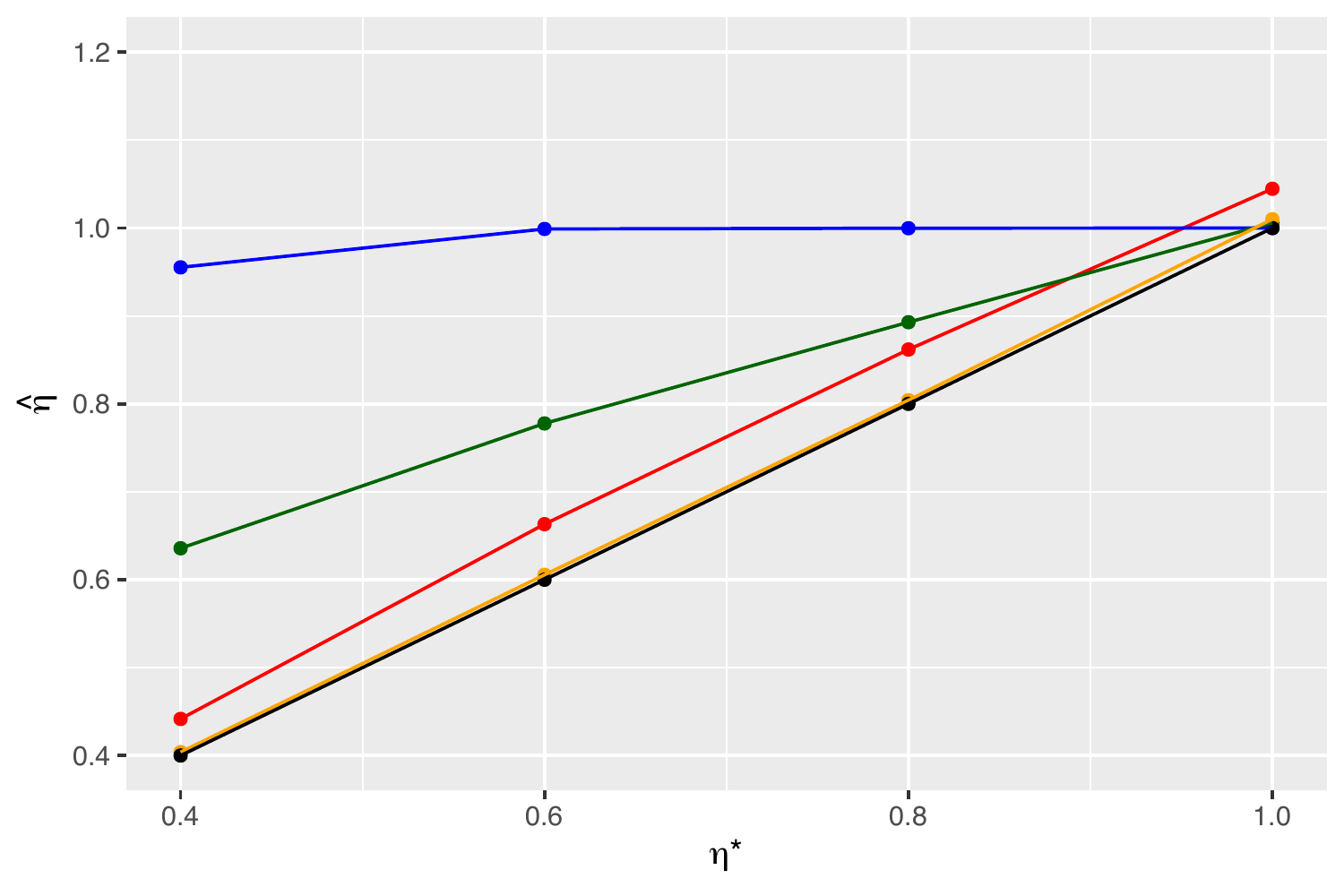}}}
\subfigure[$n=800$]{\scalebox{0.45}{\includegraphics{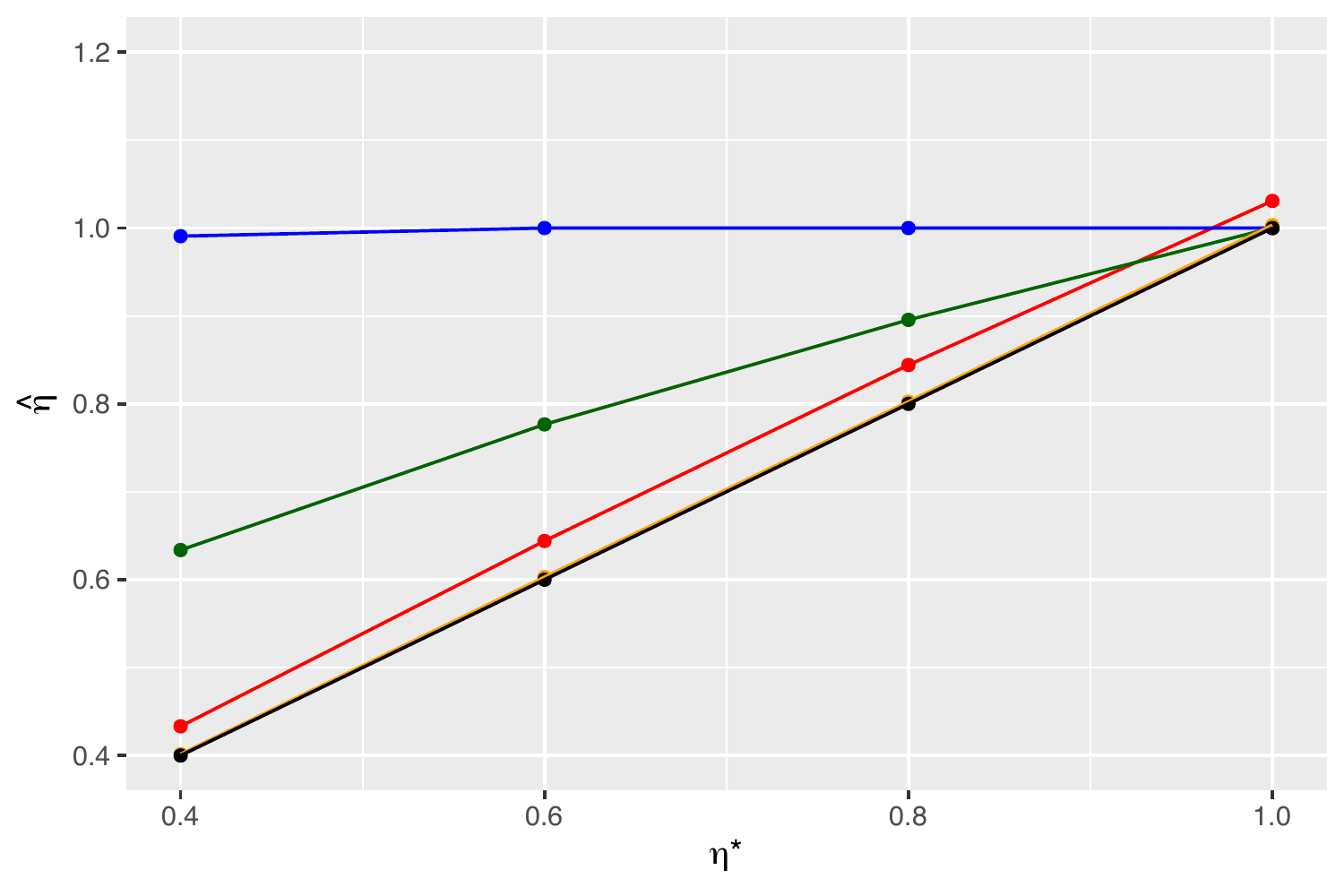}}}
\end{center}
   \caption{Average learning rate $\eta$, across 1000 replications, versus the optimal $\eta^\star = (\sigma/\sigma^\star)^2$, the closer to the diagonal line the better. True (black), GPC (red), R-SafeBayes (blue), Holmes and Walker (green), Lyddon et al.~(orange).}
   \label{f:learningrate_toyexample}
\end{figure}

\begin{figure}[t]
\begin{center}
\subfigure[$n=100$]{\scalebox{0.45}{\includegraphics{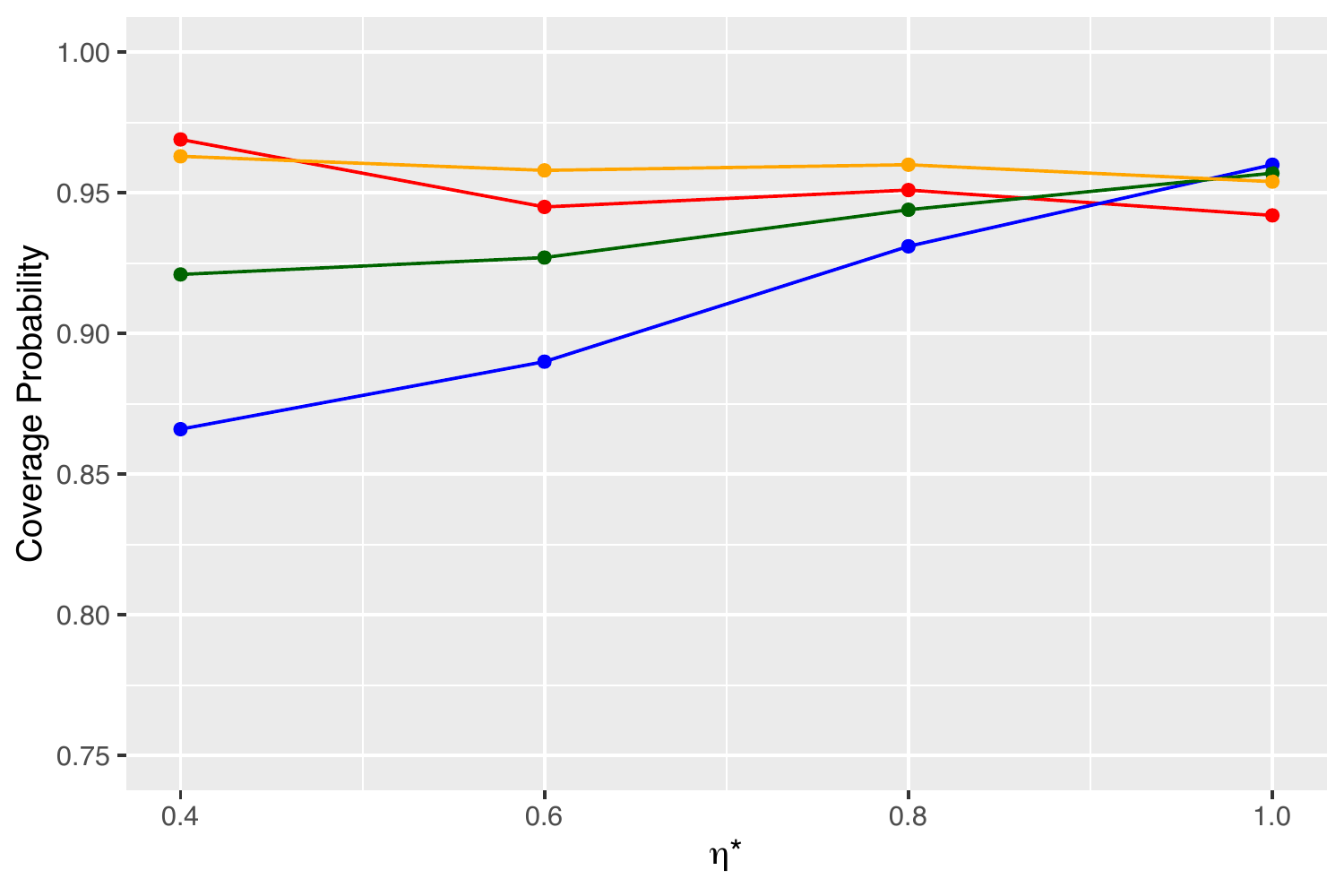}}}
\subfigure[$n=200$]{\scalebox{0.45}{\includegraphics{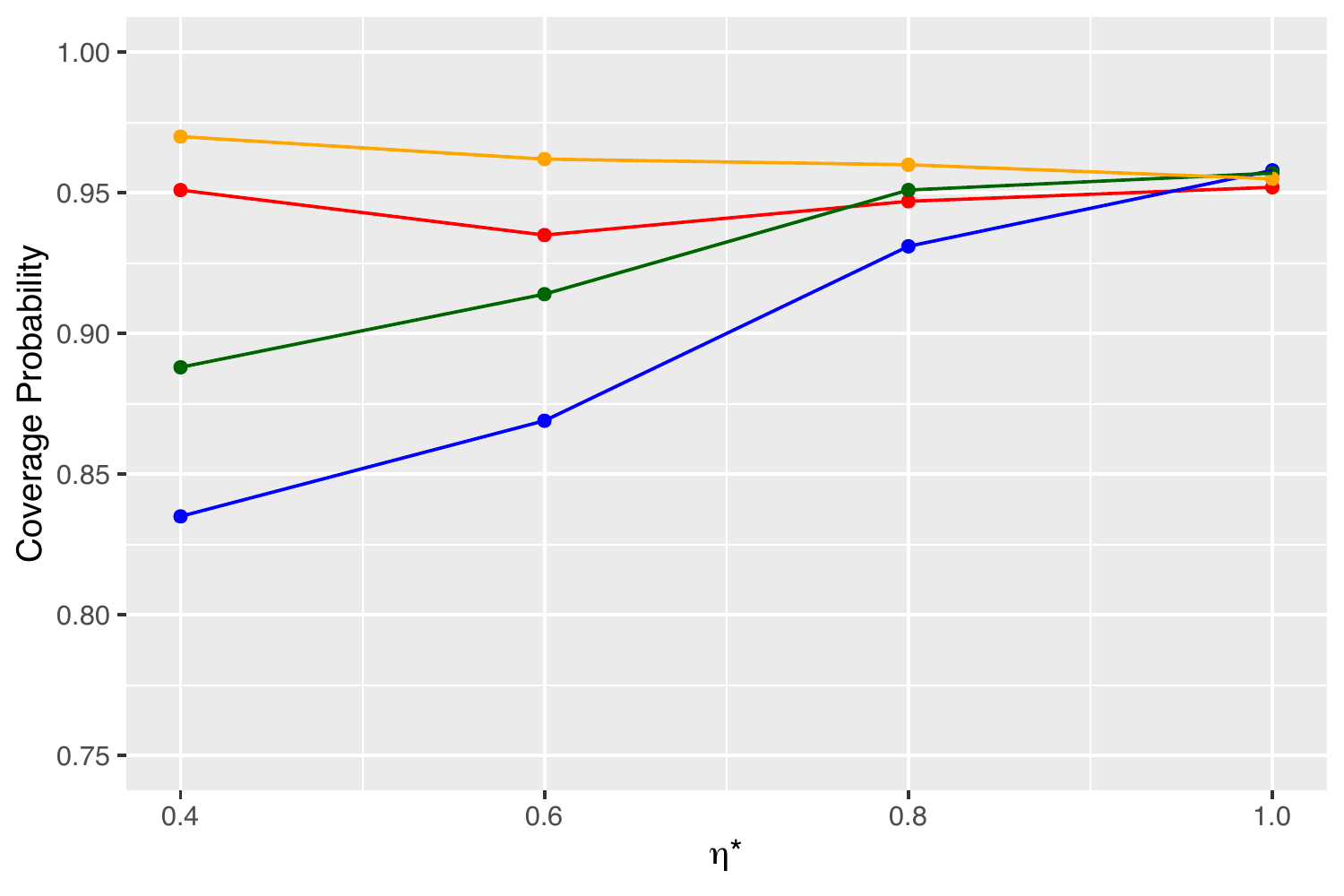}}}
\subfigure[$n=400$]{\scalebox{0.45}{\includegraphics{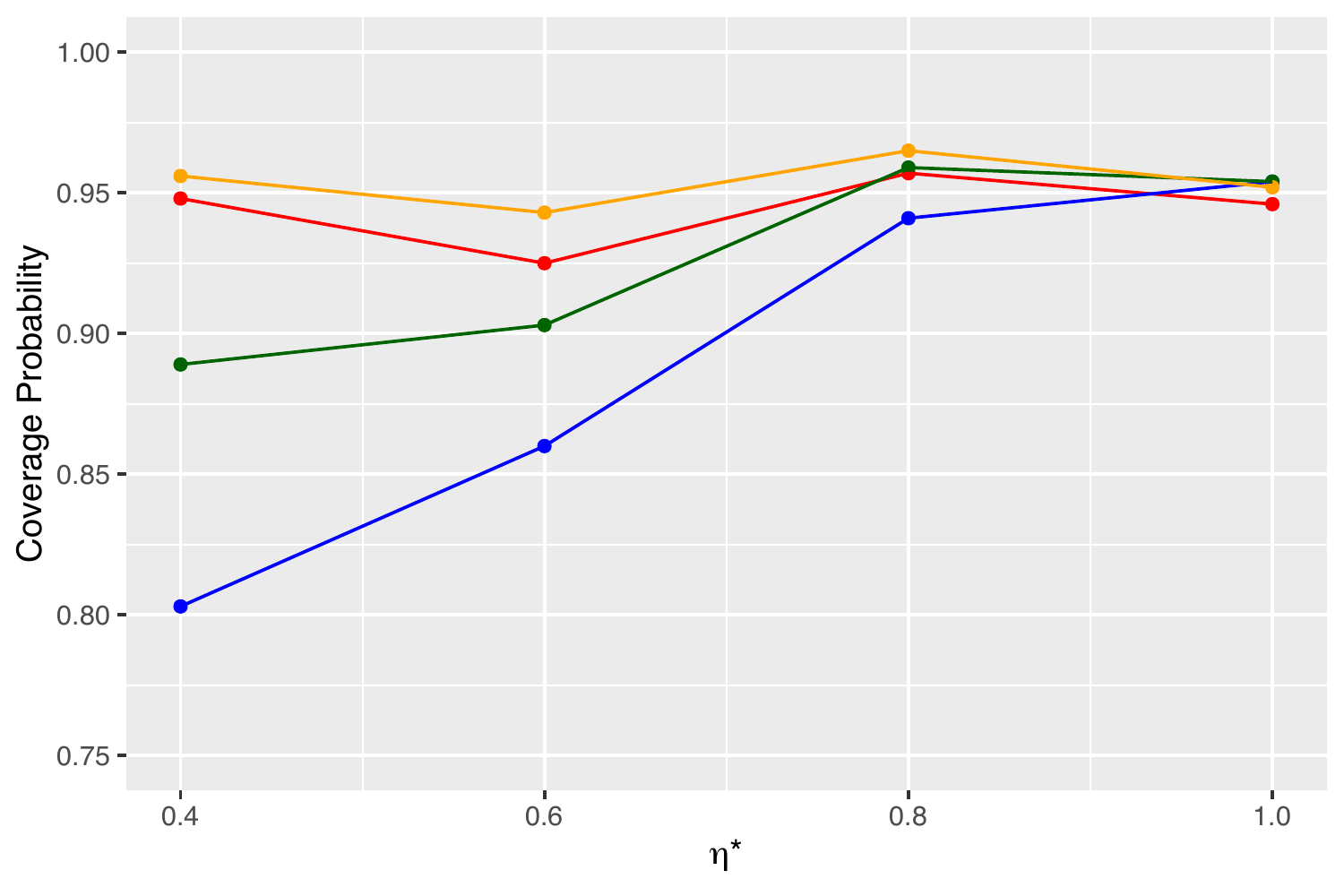}}}
\subfigure[$n=800$]{\scalebox{0.45}{\includegraphics{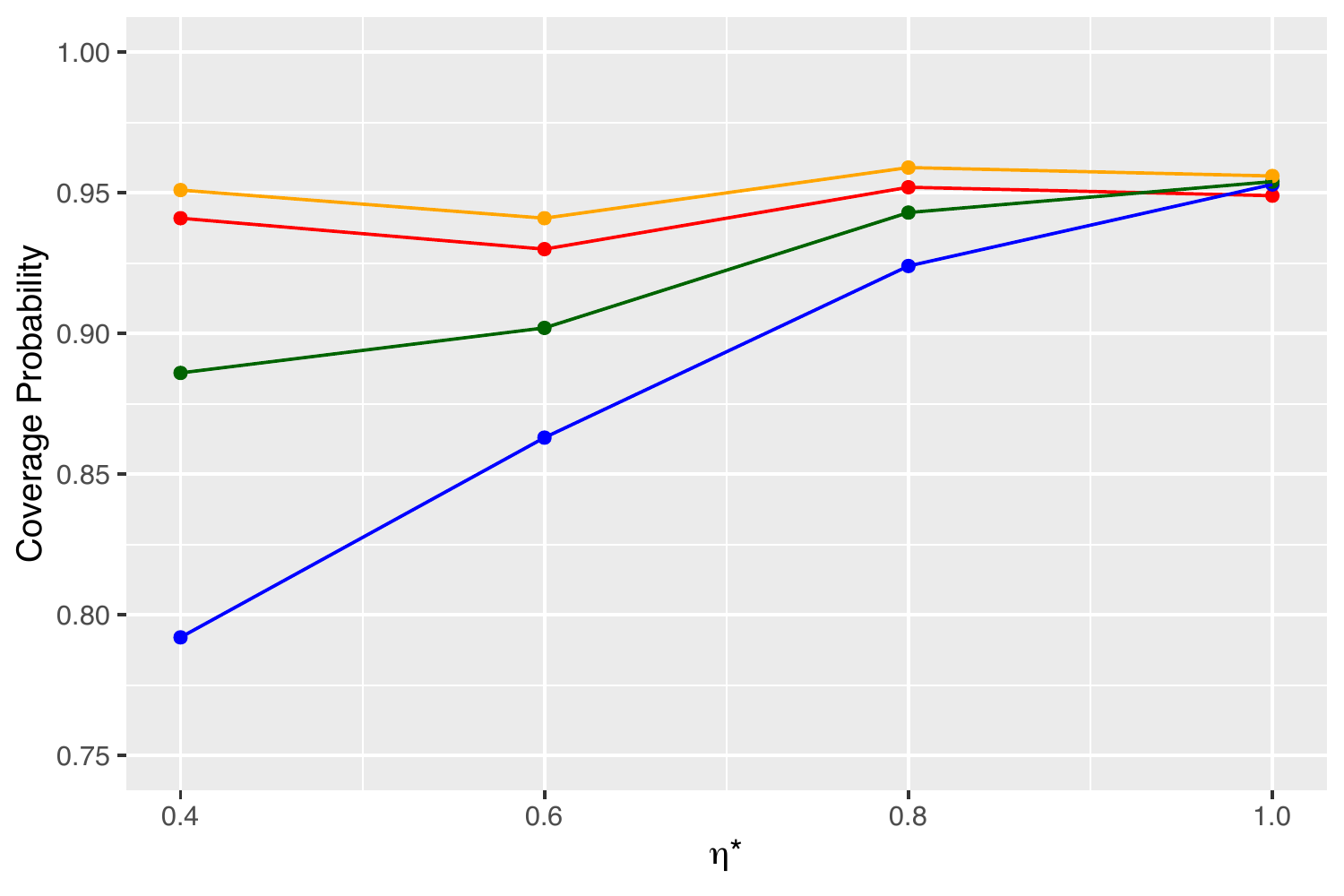}}}
\end{center}
     
   \caption{Average coverage probability of the nominal $95\%$ generalized Bayes credible intervals, across 1000 replications. GPC (red), SafeBayes (blue), Holmes and Walker (green), Lyddon et al.~(orange).}
   \label{f:coverage_toyexample}
\end{figure}

To evaluate the performance of the different learning rate methods, we simulate $1000$ data sets, for each of several different sample sizes $n$ and values of $\eta^\star = (\sigma/\sigma^\star)^2$, and compare the average estimated learning rate against $\eta^\star$; see Figure~\ref{f:learningrate_toyexample}.  If the estimated $\eta$ is close to the diagonal line $\eta=\eta^\star$, then the generalized Bayesian credible sets have coverage probability near the nominal level.  To confirm this, see  Figure~\ref{f:coverage_toyexample}.  When the degree of misspecification is relatively mild, all the methods perform well.  As the misspecification degree increases, however, we find that SafeBayes and the Holmes and Walker method have decreasing coverage probability, quickly falling below any reasonable tolerance.  On the other hand, both the Lyddon et al.~and Syring and Martin methods are able to achieve the target 95\% coverage probability over the entire range of settings.  


\section{Learning rates in linear regression}
\label{S:linear}

\subsection{Model setup}

Consider a linear regression model of the form 
\begin{equation}
\label{eq:lm}
y_i = x_i^\top \beta + \sigma \eps_i, \quad i=1,\ldots,n, 
\end{equation}
where the pairs $(x_1,y_1),\ldots,(x_n,y_n)$, taking values in $\RR^p \times \RR$, are independent, $\beta \in \RR^p$ is an unknown vector of coefficients, $\sigma$ is an unknown scale parameter, and  $\eps_1,\ldots,\eps_n$ are random error terms.  As is most common, here we will consider a model that assumes the errors $\eps_1,\ldots,\eps_n$ are iid $\nm(0,1)$, independent of $x_1,\ldots,x_n$.  If it happens that the true distribution is different from this posited model\, then, in general, we can expect an ordinary Bayes posterior to suffer from misspecification bias.  The goal here is to investigate how the different learning rate methods can help the generalized Bayes posterior to correct for this misspecification bias.  

The two key assumptions behind the textbook linear regression model are that the errors are (a) independent of covariates and (b) normally distributed.  Here we present results for two types of misspecification, namely, {\em Dependent Errors} and {\em Non-normal Errors}.  The specific form and degree of these misspecifications will be described in the following subsections.  For the comparison, the metrics we consider are 
\begin{itemize}
\item mean value of the learning rate estimates, $\hat\eta$;
\vspace{-2mm}
\item coverage probability of the $\hat\eta$-generalized Bayes 95\% highest posterior density credible sets for the full $\beta$ vector;
\vspace{-2mm}
\item mean square error of the $\hat\eta$-generalized Bayes posterior mean of $\beta$;
\vspace{-2mm}
\item the average of the marginal $\hat\eta$-generalized Bayes posterior variances for each coordinate of $\beta$. 
\end{itemize}
We are specifically interested in the learning rate and its effect on the coverage probability of the generalized posterior credible sets, so the first two metrics are clear.  The mean square error of the generalized posterior mean acts like an overall measure of bias, i.e., how far does the center of the posterior tend to be from the true parameter values.  In the examples that follow, we find that the mean square error does not vary much relative to the learning rate selection method, which confirms our intuition that the learning rate really only impacts the posterior spread.  The fourth metric is an overall measure of the spread of the generalized Bayes posterior, and we expect that those learning rate selection methods whose credible regions tend to under-cover will have smaller total variance.

\subsection{Dependent errors}
\label{SS:linear.dependent}


In the linear regression model described above, we sample the predictors $x_i$ independently from a multivariate normal distribution with mean zero, unit variance, and a first-order autocorrelation structure, i.e., $\E(x_{ij} x_{ik}) = \rho^{|j-k|}$, with correlation $\rho=0.2$.  We also set the true coefficient vector to be $\beta=(1,1,2,-1)^\top$.  The errors $\eps_1,\ldots,\eps_n$ are independent standard normal, but the error standard deviation is misspecified in the sense that the presumed constant $\sigma$ is not a constant, in particular, its value depends on the individual $x_i$ in the following way.  Let $\hat\xi_{0.05}$ and $\hat\xi_{0.95}$ denote the sample $5^\text{th}$ and $95^\text{th}$ percentiles of $x_{11},\ldots,x_{n1}$.  Then define the case-specific standard deviation as 
\[ \sigma_i = \begin{cases} s_{\text{small}} & \text{if $x_{i1} < \hat\xi_{0.05}$} \\ s_{\text{mod}} & \text{if $\hat\xi_{0.05} \leq x_{i1} \leq \hat\xi_{0.95}$} \\ 1 & \text{if $x_{i1} > \hat\xi_{0.95}$}, \end{cases} \]
where the small and moderate values, $s_\text{small}$ and $s_\text{mod}$, control the degree of the departures from constant variance.  We consider three different degrees of misspecification.  

\begin{description}
  \item[\sc Degree 1.] $s_\text{small}=0.25$ and $s_\text{mod}=0.50$;
  \vspace{-2mm}
  \item[\sc Degree 2.] $s_\text{small}=0.05$ and $s_\text{mod}=0.25$;
  \vspace{-2mm}
  \item[\sc Degree 3.] $s_\text{small}=0.01$ and $s_\text{mod}=0.10$.
\end{description}


A summary of the different learning rate selection procedures, across the different misspecification degrees and sample sizes $n \in \{100, 200, 400\}$, is presented in Table~\ref{t:two}, based on 1000 data sets for each combination. In Degree~1, where the misspecification is relatively mild, we see that all four learning rate selections perform well and similarly in terms of both the learning rates chosen---all near 1---and in the coverage probabilities.  As expected, however, as the misspecification gets more severe, in Degrees~2 and 3, the more disparity we see between the selected learning rates and, in turn, in the coverage probabilities.  Only GPC is able to achieve the nominal coverage probability in the more severe misspecification settings, while the performance of other methods can be quite poor, especially under Degree~3 with small sample sizes.  The mean square errors are more or less the same for the methods within each sample size--degree combination; and the fact that these values are small indicates the the posterior is generally centered around the target $\beta$ values.  As for the posterior spread, there is not much difference between the results in the Degree~1 case with only mild misspecification.  However, in Degrees~2 and 3, where the misspecification is more severe, we see greater difference in the posterior variance.  As expected, those methods whose posterior variance tends to be small are those who tend to have credible sets that under-cover, in many cases severely.



\begin{table}[t]
\centering
\begin{tabular}{cccccccc}
 \toprule
  Degree & $n$ & Method  & $\hat\eta$ & Coverage & MSE & Variance \\
 \toprule 
  1 & $100$ & GPC  & $0.95$ & $0.95$ & $0.05$ & $0.012$\\
  &  & SafeBayes  & $0.92$ & $0.94$ & $0.05$ & $0.014$\\
  &  & Holmes and Walker  & $1.00$ & $0.93$ & $0.05$ & $0.011$ \\
  &  & Lyddon et al.  & $1.18$ & $0.89$ & $0.05$ & $0.010$\\
\hline
  & $200$ & GPC  & $0.95$ & $0.93$ & $0.02$ & $0.006$\\
  &  & SafeBayes  & $0.92$ & $0.93$ & $0.02$ & $0.007$\\
  &  & Holmes and Walker  & $0.99$ & $0.92$ & $0.02$ & $0.006$ \\
  &  & Lyddon et al.  & $1.06$ & $0.90$ & $0.02$ & $0.005$\\
\hline
  & $400$ & GPC  & $0.94$ & $0.95$ & $0.01$ & $0.003$ \\
  &  & SafeBayes  & $0.93$ & $0.94$ & $0.01$ & $0.003$\\
  &  & Holmes and Walker  & $0.99$ & $0.94$ & $0.01$  & $0.003$\\
  &  & Lyddon et al.  & $0.99$ & $0.94$ & $0.01$ & $0.003$\\
  \toprule
2 & $100$ & GPC  & $0.79$ & $0.95$ & $0.06$ & $0.015$ \\
  &  & SafeBayes  & $0.90$ & $0.90$ & $0.06$ & $0.014$\\
  &  & Holmes and Walker  & $0.98$ & $0.89$ & $0.06$ & $0.012$  \\
  &  & Lyddon et al.  & $1.33$ & $0.76$ & $0.06$ & $0.009$ \\
\hline
  & $200$ & GPC  & $0.75$ & $0.95$ & $0.03$ & $0.008$ \\
  &  & SafeBayes  & $0.92$ & $0.90$ & $0.03$ & $0.006$\\
  &  & Holmes and Walker  & $0.97$ & $0.89$ & $0.03$ & $0.006$ \\
  &  & Lyddon et al.  & $1.11$ & $0.84$ & $0.03$ & $0.005$\\
\hline
  & $400$ & GPC  & $0.74$ & $0.94$ & $0.01$ & $0.004$\\
  &  & SafeBayes  & $0.93$ & $0.89$ & $0.01$ & $0.003$ \\
  &  & Holmes and Walker  & $0.96$ & $0.88$ & $0.01$  & $0.003$\\
  &  & Lyddon et al.  & $0.97$ & $0.88$ & $0.01$ & $0.003$\\
 \toprule
3 & $100$ & GPC  & $0.54$ & $0.98$ & $0.07$ & $0.023$\\
  &  & SafeBayes  & $0.75$ & $0.87$ & $0.07$  & $0.018$\\
  &  & Holmes and Walker  & $0.94$ & $0.80$ & $0.07$ & $0.012$ \\
  &  & Lyddon et al.  & $2.45$ & $0.38$ & $0.07$ & $0.005$\\
\hline
  & $200$ & GPC  & $0.53$ & $0.95$ & $0.04$ & $0.011$ \\
  &  & SafeBayes  & $0.76$ & $0.86$ & $0.04$ & $0.008$\\
  &  & Holmes and Walker  & $0.91$ & $0.79$ & $0.04$  & $0.006$\\
  &  & Lyddon et al.  & $1.74$ & $0.53$ & $0.04$ & $0.003$\\
\hline
  & $400$ & GPC  & $0.53$ & $0.95$ & $0.02$ & $0.005$\\
  &  & SafeBayes  & $0.78$ & $0.84$ & $0.02$ & $0.004$\\
  &  & Holmes and Walker  & $0.89$ & $0.81$ & $0.02$  & $0.003$\\
  &  & Lyddon et al.  & $1.25$ & $0.69$ & $0.02$ & $0.002$\\
  \bottomrule
\end{tabular}
\caption{Comparison of average learning rate estimates ($\hat\eta$), estimated coverage probabilities (Coverage), mean square error (MSE), and total posterior variance (Variance) across different sample sizes and misspecification degrees in the {\em Dependent Errors} example.} 
\label{t:two}
\end{table}

\subsection{Non-normal errors}
\label{SS:linear.nonnormal}

Next, we consider departures from the specified model in terms of the distribution of the error terms.  It turns out that the performance of the learning rate selection methods was less sensitive to departures from normality compared to departures from the constant-error-variance assumption.  Here we present the results for only one kind of departure from normality, namely, with heavy-tailed errors.  In particular, consider errors $\eps_1,\ldots,\eps_n$ iid from a Student-t distribution with degrees of freedom $\nu$.  As before, we consider three degrees of misspecification, each sufficiently light-tailed that the variance exists.

\begin{description}
\item[\sc Degree  1.] $\nu=5$; 
\vspace{-2mm}
\item[\sc Degree  2.] $\nu=4$; 
\vspace{-2mm}
\item[\sc Degree  3.] $\nu=3$.
\end{description}

Table~\ref{t:five} summarizes the results just like in the previous subsection.  Here, however, the differences in performance across different learning rate selection methods, sample sizes, and misspecification degrees is much smaller.  Overall the methods return similar learning rate estimates and hit the target coverage probability on the mark.  The method of Lyddon et al.~tends to select a learning rate that is too large, leading to under-coverage, but its performance tends to improve as the sample size increases.

\begin{table}
\begin{center}
\begin{tabular}{cccccccc}
 \toprule
  Degree & $n$ & Method  & $\hat\eta$ & Coverage & MSE & Variance \\
 \toprule
  1 & $100$ & GPC  & $0.98$ & $0.95$ & $0.07$ & $0.019$ \\
  &  & SafeBayes  & $0.90$ & $0.95$ & $0.07$ & $0.023$\\
  &  & Holmes and Walker  & $1.00$ & $0.94$ & $0.07$ & $0.019$  \\
  &  & Lyddon et al.  & $1.28$ & $0.87$ & $0.07$ & $0.014$ \\
\hline
  & $200$ & GPC  & $0.99$ & $0.96$ & $0.04$ & $0.009$ \\
  &  & SafeBayes  & $0.90$ & $0.96$ & $0.04$ & $0.011$\\
  &  & Holmes and Walker  & $0.98$ & $0.96$ & $0.04$  & $0.009$\\
  &  & Lyddon et al.  & $1.15$ & $0.92$ & $0.04$ & $0.008$\\
\hline
  & $400$ & GPC  & $1.00$ & $0.95$ & $0.02$ & $0.005$ \\
  &  & SafeBayes  & $0.92$ & $0.95$ & $0.02$ & $0.005$ \\
  &  & Holmes and Walker  & $0.98$ & $0.95$ & $0.02$  & $0.005$\\
  &  & Lyddon et al.  & $1.07$ & $0.94$ & $0.02$ & $0.004$ \\
  \toprule 
2 & $100$ & GPC  & $0.97$ & $0.96$ & $0.08$ & $0.024$\\
  &  & SafeBayes  & $0.89$ & $0.96$ & $0.08$ & $0.028$ \\
  &  & Holmes and Walker  & $0.99$ & $0.96$ & $0.08$ & $0.023$  \\
  &  & Lyddon et al.  & $1.38$ & $0.87$ & $0.08$ & $0.016$ \\
\hline
  & $200$ & GPC  & $0.98$ & $0.96$ & $0.04$ & $0.011$\\
  &  & SafeBayes  & $0.91$ & $0.96$ & $0.04$ & $0.013$ \\
  &  & Holmes and Walker  & $0.97$ & $0.96$ & $0.04$ & $0.011$  \\
  &  & Lyddon et al.  & $1.20$ & $0.91$ & $0.04$ & $0.009$ \\
\hline
  & $400$ & GPC  & $0.99$ & $0.96$ & $0.02$ & $0.006$\\
  &  & SafeBayes  & $0.92$ & $0.96$ & $0.02$ & $0.007$ \\
  &  & Holmes and Walker  & $0.96$ & $0.96$ & $0.02$ & $0.006$  \\
  &  & Lyddon et al.  & $1.12$ & $0.92$ & $0.02$ & $0.005$ \\
  \toprule
3 & $100$ & GPC  & $0.92$ & $0.97$ & $0.13$ & $0.044$ \\
  &  & SafeBayes  & $0.87$ & $0.97$ & $0.13$ & $0.047$ \\
  &  & Holmes and Walker  & $0.93$ & $0.96$ & $0.13$ & $0.041$  \\
  &  & Lyddon et al. & $1.69$ & $0.81$ & $0.13$ & $0.020$ \\
\hline
  & $200$ & GPC  & $0.96$ & $0.97$ & $0.06$ & $0.018$ \\
  &  & SafeBayes  & $0.90$ & $0.96$ & $0.06$ & $0.021$ \\
  &  & Holmes and Walker  & $0.92$ & $0.96$ & $0.06$ & $0.020$ \\
  &  & Lyddon et al.  & $1.37$ & $0.86$ & $0.06$ & $0.011$\\
\hline
  & $400$ & GPC  & $0.97$ & $0.96$ & $0.03$ & $0.009$ \\
  &  & SafeBayes  & $0.91$ & $0.96$ & $0.03$ & $0.010$ \\
  &  & Holmes and Walker  & $0.86$ & $0.96$ & $0.03$ & $0.012$  \\
  &  & Lyddon et al.  & $1.26$ & $0.89$ & $0.03$ & $0.006$ \\
  \bottomrule 
\end{tabular}
\end{center}
\caption{Comparison of average learning rate estimates ($\hat\eta$), estimated coverage probabilities (Coverage), mean square error (MSE), and total posterior variance (Variance) across different sample sizes and misspecification degrees in the {\em Non-normal Errors} example.}
\label{t:five}
\end{table}


\subsection{Other experiments}

Finally, we considered other types of misspecification in addition to those presented above.  These results are not presented here because all four learning rate selection methods performed similarly and displaying a table of similar numbers is not a good use of space.  But it is worth mentioning in what cases these methods perform comparably, and below is a brief summary of our findings.
\begin{itemize}
\item In cases where the heteroscedasticity is less extreme than in Section~\ref{SS:linear.dependent} above, in particular, with errors having non-constant variance but independent of $x$, we found that all four learning rate selection methods performed well.  That is, the learning rate estimates were all similar and the credible regions all had coverage probability near the nominal 95\% level. 
\vspace{-2mm}
\item The example in Section~\ref{SS:linear.nonnormal} considered heavy-tailed error distributions.  We also considered cases where the error distribution was asymmetric, e.g., skew-normal \citep{perez2018bayesian}.  Apparently, misspecification in the shape of the error distribution has little effect because, as above, the learning rate selection methods all performed well in these cases.   
\end{itemize}

\section{Learning rates in logistic regression}
\label{S:logistic}

\subsection{Model setup}

An important problem in medical statistics is estimation of the so-called {\em minimum clinically important difference} (MCID) that assesses the practical as opposed to statistical significance of a treatment.  In words, the MCID is the threshold on the diagnostic measure scale such that improvements beyond that level are associated with patients feeling better after the treatment; see, e.g., \citet{hedayat2015minimum} and the references therein.  
To set the scene, let $X \in \RR$ denote the patient's diagnostic measure, e.g., the pre-treatment minus post-treatment difference in blood pressure, and let $Y \in \{-1,+1\}$ denote the patient-reported indicator of whether they felt the treatment was effective, with ``$y=+1$'' indicating effective.  The quantity of interest, $\theta$, the MCID, is the cutoff on the $X$ scale such that the indicator $1\{X > \theta\}$ is most highly associated with $Y$. More precisely, the MCID is defined as
\[ \theta = \arg \min_\vartheta P\{Y \neq \text{sign}(X-\vartheta)\}, \]
where $\text{sign}(0)=1$. Clearly, $\theta$ depends on the unknown joint distribution $P$ of $(X,Y)$.

Towards inference on the MCID, it is natural to introduce a statistical model for $P$.  It would be difficult to develop a model for which $\theta$ is directly a model parameter, but one idea would be to use a logistic regression model with $Y$ as the binary response and $X$ as a continuous predictor.  That is, the logistic regression model states that 
\[ (Y \mid X=x) \sim \rad\bigl( F(\beta_0 + \beta_1 x) \bigr), \]
where $\rad(p)$ denotes a Rademacher distribution, i.e., a binary distribution on $\{-1,+1\}$, with probability mass $p$ assigned to the value $+1$, and $F$ is a logistic distribution function with $F(u) = \{1 + e^{-u}\}^{-1}$, for $u \in \RR$.  The logistic regression model is determined by the unknown parameters $(\beta_0,\beta_1)$.  Since the MCID $\theta$ also satisfies $P(Y = +1 \mid X=\theta) = \tfrac12$, if the above model is assumed, then 
\[ \theta = -\beta_0 / \beta_1. \]
Given independent observations $(X_1,Y_1),\ldots,(X_n,Y_n)$ from this model, a posterior distribution for $(\beta_0,\beta_1)$, generalized Bayes or otherwise, can be obtained.  From this, one can readily obtain the corresponding posterior distribution of $\theta$ via the identity above.  

Of course, this model could easily be misspecified.  So it is of interest to investigate what happens with the the generalized Bayes posterior with suitably chosen learning rates when the logistic link function $F$ is incorrectly specified.

\subsection{Results}

We fit a misspecified logistic regression model, i.e., where the diagnostic measure $X$ comes from a normal mixture model with distribution function 
\[ F^\star(x)=0.7 \, \Phi(x \mid 5, 1) + 0.3 \, \Phi(x \mid \mu, 1), \]
and the patient reported effectiveness indicator is $(Y \mid X=x) \sim \rad(F^\star(x))$.  Here the quantity $\mu$ controls the degree of misspecification, with $\mu$ closer to 5 corresponding to ``less misspecification''---relative to the logistic link function $F$ above---compared to $\mu$ further from 5.  The three specific degrees considered are:
\begin{description}
  \item[\sc Degree 1.] $\mu=7$;
  \vspace{-2mm}
  \item[\sc Degree 2.] $\mu=8$;
  \vspace{-2mm}
  \item[\sc Degree 3.] $\mu=9$.
\end{description}

For the posited logistic regression model, we follow \citet[][Example~7.11]{casella} and take a default prior distribution for $(\beta_0, \beta_1)$ to be  
\[ \pi(\beta_0,\beta_1)=\hat{b}^{-1} \exp\{\beta_0 - \hat b^{-1} e^{\beta_0} \}, \]
which is simply a flat prior for $\beta_1$ and an exponential prior for $e^{\beta_0}$ with scale $\hat b = \exp(\hat\beta_0 + \gamma)$, where $\hat\beta_0$ is the maximum likelihood estimator and $\gamma \approx 0.5772$ is Euler's constant.  This default prior is used here, rather than, say, the P\'olya--gamma prior of \citet{polson.scott.windle.2013}, because embedding its more sophisticated posterior sampling scheme inside the learning rate selection procedures was simply too expensive. 

The goal is, as in the previous section, to investigate the extent to which the learning rate selection methods can help the generalized Bayes posterior distribution to overcome the model misspecification, and Table~\ref{t:six} summarizes the results.  There we present the average learning rate value, the coverage probability of 95\% credible intervals for $\theta$, the average length of those credible intervals, and mean square error, all based on 500 replications, for each pair of $\mu$ and sample size $n$.  Here we see that, in the Degree~1 case where misspecification is relatively mild, the methods perform reasonably well in terms of coverage probability, but things get worse as sample size increases, a symptom of the model misspecification bias.  For the Degree~2--3 cases with even more model misspecification, all the methods perform quite poorly.  Apparently none of the learning rate selection methods can help the posterior overcome the relatively severe model misspecification bias in this example. 


\begin{table}[t]
\centering
\begin{tabular}{ccccccc}
 \toprule
  Degree & $n$ & Method   & $\hat\eta$ & Coverage & Length & MSE \\
 \toprule
1 & $100$ & GPC  & $0.904$ & $0.938$ & $0.883$ & $0.051$ \\
  &  & SafeBayes  & $0.790$ & $0.953$ & $0.987$ & $0.053$ \\
  &  & Holmes and Walker  & $0.999$ & $0.927$ & $0.834$ & $0.051$ \\
  &  & Lyddon et al.  & $1.003$ & $0.923$ & $0.830$ & $0.051$ \\
\hline
  & $200$ & GPC  & $0.977$ & $0.914$ & $0.575$ & $0.029$ \\
  &  & SafeBayes & $0.913$ & $0.916$ & $0.599$ & $0.029$\\
  &  & Holmes and Walker & $0.999$ & $0.902$ & $0.568$ & $0.030$ \\
  &  & Lyddon et al. & $1.003$ & $0.923$ & $0.830$ & $0.030$ \\
\hline
  & $400$ & GPC & $0.910$  & $0.912$ & $0.418$ & $0.015$\\
  &  & SafeBayes  & $0.822$ & $0.926$ & $0.450$ & $0.015$ \\
  &  & Holmes and Walker& $0.999$ & $0.890$ & $0.397$ & $0.015$ \\
  &  & Lyddon et al. & $0.988$ & $0.892$ & $0.401$ & $0.015$\\
  \toprule
2 & $100$ & GPC  & $0.786$ & $0.866$ & $1.148$  &  $0.139$\\
  &  & SafeBayes & $0.890$ &$0.893$ & $1.283$ &$0.138$ \\
  &  & Holmes and Walker & $1.000$ & $0.836$ & $1.071$  &  $0.137$\\
  &  & Lyddon et al.& $0.986$ & $0.838$ & $1.085$  & $0.137$\\
\hline
  & $200$ & GPC  & $0.970$ & $0.788$ & $0.752$ & $0.083$\\
  &  & SafeBayes  & $0.906$ & $0.816$ & $0.786$ & $0.082$\\
  &  & Holmes and Walker  & $1.001$ & $0.776$ & $0.741$ & $0.082$ \\
  &  & Lyddon et al.& $0.974$ & $0.727$ & $0.742$ & $0.087$\\
\hline
  & $400$ & GPC  & $0.901$ & $0.622$ & $0.539$  &  $0.067$\\
  &  & SafeBayes & $0.833$  & $0.646$ & $0.574$ & $0.067$\\
  &  & Holmes and Walker& $0.999$ & $0.564$ & $0.511$ & $0.067$ \\
  &  & Lyddon et al. & $0.969$ & $0.588$ & $0.518$ & $0.067$\\
  \toprule
3 & $100$ & GPC  & $0.955$ & $0.742$ & $1.412$ & $0.345$ \\
  &  & SafeBayes & $0.891$ & $0.750$ & $1.482$ & $0.346$ \\
  &  & Holmes and Walker & $1.001$ & $0.726$ & $1.377$ & $0.344$  \\
  &  & Lyddon et al.  & $0.970$ & $0.731$ & $1.385$ & $0.343$ \\
\hline
  & $200$ & GPC  & $0.951$ & $0.532$ & $0.976$ & $0.266$ \\
  &  & SafeBayes & $0.887$ & $0.562$ & $1.024$ &  $0.264$\\
  &  & Holmes and Walker & $1.004$ & $0.502$ & $0.946$& $0.265$\\
  &  & Lyddon et al. & $0.955$ & $0.510$ & $0.963$ & $0.261$ \\
\hline
  & $400$ & GPC & $0.820$ & $0.244$ & $0.693$ &  $0.247$\\
  &  & SafeBayes & $0.893$ & $0.290$ & $0.743$& $0.248$\\
  &  & Holmes and Walker & $1.000$  & $0.228$ & $0.657$ & $0.248$ \\
  &  & Lyddon et al. & $0.951$ & $0.236$ & $0.671$ & $0.247$\\
  \bottomrule
\end{tabular}
\caption {Summary of learning rate selection method performance in the misspecified logistic regression example based on 500 replications.}
\label{t:six}
\end{table}


\subsection{A Gibbs posterior} 

The generalized Bayes posterior is not able to overcome this apparently rather severe form of misspecification bias.  As an alternative, we can consider a different type of posterior construction, the so-called {\em Gibbs posterior}.  What distinguishes a Gibbs from a generalized Bayes posterior is that the former is defined by a loss function while the latter is defined by a likelihood.  In the present context, the Gibbs posterior is a more appropriate approach since the MCID is not naturally defined as a model parameter in a likelihood.  Therefore, it is possible to {\em directly} define the Gibbs posterior for the MCID, $\theta$, as opposed to indirectly through a likelihood for $\beta$ and marginalizing to $\theta$.  

Define the loss function $\ell_\theta(x,y) = \frac12 \{1 - y \, \text{sign}(x-\theta)\}$ and the corresponding risk (expected loss) $R(\vartheta) = P \ell_\vartheta$.  As \citet{hedayat2015minimum}, showed, the MCID is the minimizer of $R$, i.e., $\theta^\star = \arg\min_\theta R(\theta)$. So the goal is to construct an empirical version of the risk function, and then a sort of posterior distribution that will concentrate around values that make the empirical risk small.  For the empirical risk, let 
\[ R_n(\theta) = \frac1n \sum_{i=1}^n \ell_\theta(X_i, Y_i). \]
Then the Gibbs posterior distribution for $\theta$ has a density function defined as 
\[ \pi_n^{(\eta)}(\theta) \propto e^{-\eta n R_n(\theta)} \, \pi(\theta), \]
where $\eta > 0$ is, as before, the learning rate.  In principle, all the different learning rate selection methods considered above can be applied to the Gibbs posterior framework to choose an appropriate value of $\eta$.  Here, however, the loss function is not differentiable, which creates a challenge for the methods of \citet{holmes2017assigning} and \citet{lyddon2019general}.  Therefore, in what follows, we only compare GPC and SafeBayes.  

The results in Table \ref{t:seven} are to illustrate the performance of finding the learning rate using Gibbs posterior with scaling algorithm in \cite{syring2019calibrating} and \cite{grunwald.safe}. Since the true MCID is almost certain to be in the range of observed $X$ values, the results here for both GPC and SafeBayes are based on uniform prior on $[X_{(1)}, X_{(n)}]$, the sample range.  
Here we observe that the GPC is able to choose the learning rate such that the desired 95\% coverage target for each sample size.  SafeBayes, on the other hand, tends to choose too large of a learning rate, leading to (sometimes severe) under-coverage.


\begin{table}[t]
\centering
\begin{tabular}{ccccccc}
 \toprule
  Degree & $n$ & Method & $\hat\eta$ & Coverage & Length & MSE\\
 \toprule
1 & $100$ & GPC & $0.496$ & $0.948$ & $1.579$ & $0.103$\\
  &  & SafeBayes  & $0.982$ & $0.810$ & $0.910$& $0.111$\\
\hline
  & $200$ & GPC & $0.396$ & $0.932$ & $1.177$ & $0.060$\\
  &  & SafeBayes  & $0.986$ & $0.700$ & $0.587$ & $0.073$\\
\hline
  & $400$& GPC  & $0.292$ & $0.952$ & $0.971$ & $0.033$\\
  &  & SafeBayes & $0.975$ & $0.588$ & $0.378$ & $0.048$\\
  \toprule
2 & $100$ & GPC & $0.444$ & $0.972$ & $2.251$ & $0.228$\\
  &  & SafeBayes  & $0.966$ & $0.830$ & $1.216$ & $0.201$\\
\hline
  & $200$ & GPC & $0.339$ & $0.950$ & $1.797$ & $0.143$\\
  &  & SafeBayes & $0.967$ & $0.700$ & $0.750$ & $0.123$\\
\hline
  & $400$ & GPC & $0.246$ & $0.970$ & $1.396$ & $0.073$\\
  &  & SafeBayes  & $0.964$ & $0.592$ & $0.490$  & $0.066$\\
  \toprule
 3 & $100$ & GPC  & $0.408$ & $0.966$ & $3.076$ & $0.548$\\
  &  & SafeBayes & $0.953$ & $0.804$ & $1.536$ & $0.372$\\
\hline
  & $200$ & GPC  & $0.313$ & $0.958$ & $2.452$ & $0.351$\\
  &  & SafeBayes  & $0.964$ & $0.692$ & $0.892$ & $0.205$\\
\hline
  & $400$ & GPC & $0.231$ & $0.964$ & $1.953$ & $0.187$\\
  &  & SafeBayes & $0.965$ & $0.618$ & $0.544$ & $0.080$\\
  \bottomrule
\end{tabular}
\caption {Summary of GPC and SafeBayes learning rate selection method performance using a Gibbs posterior, based on 500 replications.} 
\label{t:seven}
\end{table}

\section{Conclusion}
\label{S:discuss}

In this paper, we investigated the performance of several recently proposed procedures for choosing the learning rate parameter in generalized Bayes models.  The goal was to see which, if any of these methods, are able to overcome the model misspecification bias and give valid posterior uncertainty quantification.  While there are some models that are too severely misspecified for a simple learning rate adjustment to accommodate, but we did find that this can, in fact, be successful when misspecification is only relatively mild. 

A take-away message is that, among the learning rate selection methods considered here, the GPC algorithm of \citet{syring2019calibrating} seems to be best suited overall for calibrating the generalized Bayes credible regions.  This is not surprising, given that is precisely what the GPC algorithm is designed to do.  GPC is computationally more expensive than, say, the method of \citet{lyddon2019general}, but our results here suggest that the extra time/effort is well spent.  Although GPC has been shown to have very good empirical performance here and in a number of other references, there is still no formal proof that it does indeed provide valid posterior uncertainty quantification.  

Finally, our focus here was exclusively on inference, but it would be of interest to see if/how different learning rate selection methods might assist in generalized Bayes prediction.  After all, the prediction problem is one where it is possible to perform well even without a model, so developing a learning rate selection method that would correct for certain kinds of model misspecification, e.g., misspecified tails, should be within reach.  That is, can a suitable choice of learning rate ensure that quantiles of the posterior predictive distribution achieve the nominal prediction coverage probability?

\section*{Acknowledgments}

This work is partially supported by the U.S.~National Science Foundation, DMS--1811802.

\medskip
\bibliography{sample}
\end{document}